\documentclass[12pt,preprint]{aastex}
\usepackage{natbib}
\begin{document}

\title{Spatially Resolving the Inner Disk of TW Hya}

\author{J. A.  Eisner,\altaffilmark{1,2} E. I. Chiang,\altaffilmark{1,3}
\& L.A. Hillenbrand\altaffilmark{4}}
\altaffiltext{1}{University of California at Berkeley, 
Department of Astronomy, 601 Campbell Hall, Berkeley, CA 94720}
\altaffiltext{2}{Miller Fellow}
\altaffiltext{3}{Alfred P. Sloan Research Fellow}
\altaffiltext{4}{California Institute of Technology,  
Department of Astronomy MC 105-24, Pasadena, CA 91125}
\email{jae@astron.berkeley.edu}


\keywords{stars: planetary systems: protoplanetary 
disks---stars: pre-main-sequence---stars: individual (TW Hya)}

%

\begin{abstract}
We present Keck Interferometer observations of TW Hya that spatially
resolve its emission at 2 $\mu$m wavelength.  Analyzing
these data together with existing $K$-band veiling 
and near-infrared photometric measurements,
we conclude that the inner disk consists of optically thin,
sub-micron-sized dust extending from $\sim 4$ AU to within 0.06 AU of the 
central star.  The inner disk edge may be magnetospherically
truncated.
Even if we account for the presence of gas in the inner disk, these small
dust grains have survival times against radiation blow-out that are orders of
magnitude shorter than the age of the system,  suggesting
continual replenishment through collisions of larger bodies.
\end{abstract}

\section{Introduction}
TW Hya is a nearby \citep[$\sim 51$ pc;][]{MAMAJEK05}, young 
\citep[$\sim 10$ Myr;][]{WEBB+99} 
star surrounded by an accretion disk that evinces a large inner hole
as judged from the observed spectral energy distribution 
\citep[SED;][]{CALVET+02}.  Unusually low excess emission
at wavelengths $\lambda \la 10$ $\mu$m can be modeled with an optically thick
disk whose inner edge is located $\sim 4$ AU from the central star.  
While observations of 10 $\mu$m silicate emission \citep{SLR00,UCHIDA+04}
together with non-zero excess at 2 $\mu$m \citep{JV01} suggest the presence of 
at least some
dust grains with sizes less than a few microns at stellocentric distances
$R \la 4$ AU, this inner disk material appears optically thin,
and has been estimated to constitute less than a lunar mass \citep{CALVET+02}.
Detection of warm gas \citep{HERCZEG+04,RETTIG+04} and accretion signatures 
\citep{MUZEROLLE+00,AB02} confirm that the region inside 4 AU is not 
devoid of material.

At $R \sim 4$ AU, dust temperatures ($\sim 100$ K) are substantially 
lower than sublimation temperatures for silicate grains 
\citep[$\ga 1500$ K; e.g.,][]{POLLACK+94}.
This suggests the optically thick outer disk is truncated at 4 AU
by a mechanism
other than dust sublimation.  Large holes inferred from SEDs are commonly
attributed to planets, which may clear gaps about their orbits.
A planet impedes accretion of material outside its orbit, while inner
disk material is free to drain onto the central star
\citep{GT82,BRYDEN+99,RICE+03}.  However, a viscous outer disk causes inward 
migration of planets and their associated gaps \citep[e.g.,][]{LP86,WARD97}.  
Thus,  inner holes would be filled in on the viscous 
timescale unless the disk
is less massive than the planet, in which case the timescale is lengthened
by the mass ratio between the planet and the disk \citep{CHIANG03}.
Since the outer disk of TW Hya is massive 
\citep[$\ga 0.1$ M$_{\odot}$;][]{WEINBERGER+02,WILNER+05}, 
planets are unlikely to
preserve inner clearings over the lifetime of the system
($\sim 10$ Myr) unless the outer disk is unusually inviscid
\citep[$\alpha \la 10^{-5}$;][]{SS73}.

An alternative explanation for SED-inferred holes
is that dust grains have grown larger than a few microns, depleting
the population of small grains that would produce the near-IR
emission.  Scattered light and long-wavelength emission from the outer
($R > 10$ AU) disk of TW Hya suggest substantial grain growth, 
up to cm sizes \citep{WEINBERGER+02,WILNER+05}, supporting the hypothesis
of grain coagulation.  The small amount of sub-micron-sized dust required to 
explain emission at $\lambda \la 10$ $\mu$m and the spectral shape of the
10 $\mu$m silicate feature \citep{CALVET+02,UCHIDA+04}
may represent the tail at small sizes of a grain size distribution that
peaks at sizes much larger than a micron.  Furthermore, as we discuss below, 
small dust grains are short-lived in the TW Hya disk and demand
continual replenishment, possibly from collisions of larger parent bodies. 
The population of sub-micron-sized dust in the inner disk may not be 
primordial.

Here we present observations with the Keck Interferometer that 
spatially resolve the inner disk around TW Hya for the first time.
Previous observations at sub-millimeter to centimeter 
wavelengths spatially resolved emission at larger radii 
($> 10$ AU) and enabled powerful constraints on the 
outer disk structure and dust properties \citep{QI+04,WILNER+05}. 
With our near-IR interferometric observations, we extend this analysis to
the inner disk.  By combining spatially resolved measurements with spectral
information, we determine the radial distribution, temperature, and
approximate grain sizes of dust.  We
confirm that the inner disk is populated by small amounts of sub-micron-sized
dust, and show that the inner radius\footnote{In the remainder of
the paper, the ``inner radius'' refers to the inner edge of the
optically thin inner disk, not the inner edge of the optically
thick outer disk. We take the latter to be located at
$R \sim 4$ AU based on previous 
modeling by \citet{CALVET+02}.} 
of this optically thin disk occurs further from 
the star than previous spatially unresolved observations imply.

\section{Observations \label{sec:obs}}
\begin{figure}
\plotone{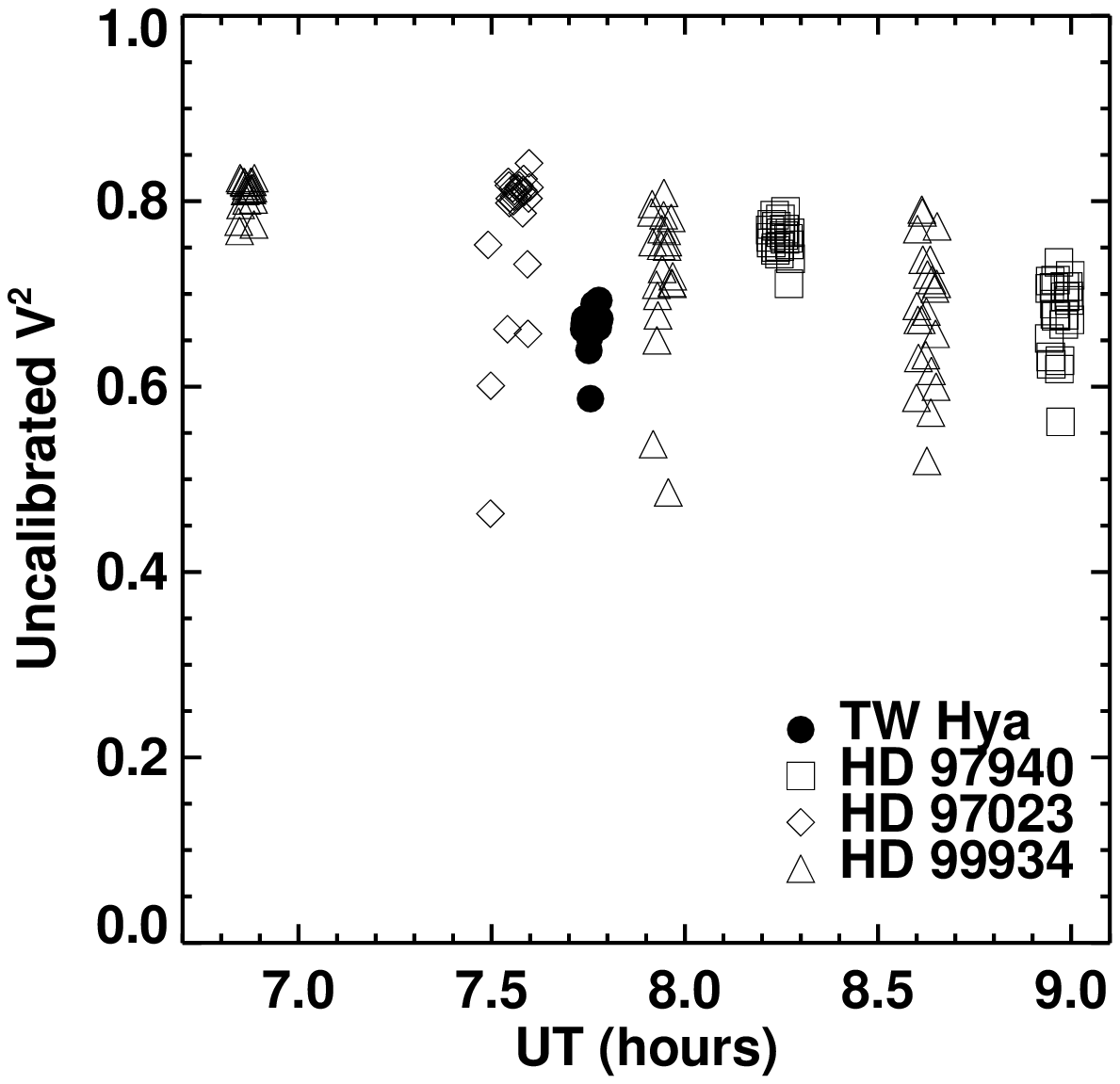}
\caption{Uncalibrated, squared visibilities ($V^2$) measured by KI for
TW Hya and three unresolved calibrator sources.  The smaller $V^2$ value
of TW Hya relative to the calibrators indicates that this source is angularly
resolved.  Scatter in the uncalibrated $V^2$ is due to a combination
of instrumental and atmospheric effects, including variable performance
of the Keck AO systems and phase jitter arising from atmospheric motions
and instrumental vibrations.
\label{fig:rawv2}}
\end{figure}

We observed TW Hya near $K$-band ($\lambda_K = 2.14$ $\mu$m; 
$\Delta \lambda =0.3$ $\mu$m) with the Keck Interferometer (KI) 
on April 21, 2005 (UT 20050421). 
The KI is a fringe-tracking  Michelson interferometer that 
combines light from the two 10-m Keck apertures
and provides an angular resolution of $\la 5$ mas \citep{CW03,COLAVITA+03}.  
A single 130s observation of TW Hya was obtained between observations of three
calibrators (HD 97023, HD 97940, and HD 99934). Figure \ref{fig:rawv2} shows
the uncalibrated visibilities of the target and calibrators.

Data calibration is described by \citet{EISNER+05}; here
we summarize the procedure.  We first determine the system
visibility (i.e., the point source response of the interferometer) from 
the weighted mean of observations of unresolved calibrators \citep{BODEN+98}. 
Source and calibrator data are corrected for 
detection biases \citep{COLAVITA99} and integrated into 5s 
blocks. The calibrated $V^2$ for the target source is averaged over all 5s 
blocks, with uncertainties given by the quadrature addition
of the internal scatter and the uncertainty in the system visibility. 
The calibrated, normalized, squared visibility measured for TW Hya at
$r_{uv}=28$ M$\lambda$ ($u=49.6$ m, $v=36.7$ m) is $V^2=0.88 \pm 0.05$.

\section{Modeling \label{sec:mod}}
We model our measured $V^2$ together with previous photometric
measurements at $\lambda=1$--5 $\mu$m \citep{WEBB+99,SLR00}, 
assuming photometric uncertainties of 10\%.
The circumstellar-to-stellar flux ratio is critical for modeling
the circumstellar component of the visibility and near-IR SED 
\citep[e.g.,][]{EISNER+04}; we utilize a previous measurement of 
this ratio at 2 $\mu$m, $r_K = 0.07 \pm 0.04$ \citep{JV01}.
Our model consists of the central star and an optically thin disk.
We model the central star using a Kurucz stellar atmosphere with 
radius $R_{\ast} =$ R$_{\odot}$, temperature 
$T_{\ast} = 4000$ K, mass $M_{\ast}=0.7$ M$_{\odot}$, surface gravity 
$\log g$ (cm s$^{-2}$) $= 4.5$,\footnote{While the computed
$\log g \approx 4.3$ for our assumed values of $R_{\ast}$ and $M_{\ast}$, 
we adopt $\log g=4.5$ since Kurucz models exist for this value.  Our results
are insensitive to this small difference.} and distance $d=51$ pc 
\citep{WEBB+99,JV01,AB02,MAMAJEK05}.  

The optically thin disk extends from an inner truncation radius, 
$R_{\rm in}$, to an outer radius, $R_{\rm out}$.
Since most of the near-IR emission is generated close to $R_{\rm in}$ where
the hottest dust resides, our results are insensitive to $R_{\rm out}$;
for simplicity, we assume $R_{\rm out}=4$ AU.
The mass surface density of the inner disk is parameterized  as
$\Sigma=\Sigma_0 (R/ {\rm AU})^{-3/2}$, where $R$ is the
stellocentric radius and $\Sigma_0$ is the surface density at $R=1$ AU.  

We assume dust grains are of a single size and
adopt a simple prescription for the frequency-dependent dust opacity,
$\kappa_{\nu} = \kappa_{K} (\nu/\nu_{K})^{\beta}$ cm$^2$ g$^{-1}$.  For
sub-micron-sized grains, we set $\beta=1$ and $\kappa_K = 10^3$.  For 
larger grains with sizes $\sim 10 \mu$m, we take $\beta=0$ and
$\kappa_K=10^2$.  These choices are compatible with previous computations
of opacities at $\lambda = 0.1$--5 $\mu$m 
by \citet{MN93}.  Our normalizations are $\sim 10^2$ times
higher than theirs since our $\kappa_\nu$ is the dust-mass opacity 
as opposed to the dust+gas opacity; i.e., the units of $\kappa_\nu$ are
cm$^2$ per g of dust.  

We compute the dust temperature under the assumption that the disk is optically
thin:
\begin{equation}
T_{\rm dust}(R) = T_{\ast} \left(\frac{R_{\ast}}{2R}\right)^{2/(4+\beta)}.
\label{eq:tr}
\end{equation}
The total flux of the inner disk 
is derived by dividing the disk into annuli,
computing the flux for each annulus, and summing the annular fluxes.  
Similarly, model visibilities are computed for each annulus, and the 
visibility for the entire inner disk is given by the flux-weighted 
sum of the annular visibilities. The flux in an annulus of infinitesimal
width $dR$ is
\begin{equation}
dF_{\nu}(R)  = \frac{2\pi}{d^2} B_{\nu} \left(T_{\rm dust} \right) 
\tau_{\nu} \: R \: dR ,
\label{eq:fr}
\end{equation}
where $B_\nu$ is the Planck function and 
$\tau_{\nu}$ is the vertical optical depth,
\begin{equation}
\tau_{\nu}(R) = \kappa_{\nu} \Sigma = \kappa_{\nu} \Sigma_0 
\left(\frac{R}{\rm AU}\right)^{-3/2}.
\label{eq:taur}
\end{equation}
The normalized visibility for an annulus extending from $R_1$
to $R_2 = R_1+dR$ is given by the difference
of visibilities for uniform disks having radii equal to
$R_1$ and $R_2$:
\begin{equation}
V(R) = \frac{\lambda d}{2\pi r_{uv} (R_2^2 - R_1^2)}
\left[R_2J_1 \left(\frac{2\pi r_{uv} R_2}{\lambda d}\right) - 
R_1J_1 \left(\frac{2\pi r_{uv} R_1}{\lambda d}\right)\right].
\end{equation}
Here $r_{uv} = 28$ M$\lambda$ is the $uv$ radius, $\lambda=\lambda_K$ is the
observing wavelength, and $J_1$ is
a first-order Bessel function.

The total flux density from the disk and the central star equals
\begin{equation}
F_{\nu, \rm tot} = F_{\nu, \ast} + 
\int_{R_{\rm in}}^{R_{\rm out}} d F_{\nu}(R).
\end{equation}
The veiling at 2 $\mu$m is
\begin{equation}
r_K = \frac{1}{F_{K, \ast}} \int_{R_{\rm in}}^{R_{\rm out}} d F_{K}(R), 
\end{equation}
where $F_K$ is the flux density at $\lambda = \lambda_K$.
The squared visibility of the model at 2 $\mu$m is
\begin{equation}
V^2 = \left[\frac{F_{K , \ast}V_{\ast} 
+ \int_{R_{\rm in}}^{R_{\rm out}} d F_{K}(R)V(R)}
{F_{K, \rm tot}}\right]^2,
\end{equation}
where $V_{\ast}=1$ is the visibility of the unresolved central star.

We solve for the best-fit parameters of the model
by computing the $\lambda=1$--5 $\mu$m fluxes, $r_K$, and $V^2$ 
for a grid of values of $R_{\rm in}$ and $\Sigma_0$, and minimizing the
$\chi^2$-residuals between model and data.
Uncertainties for best-fit parameters are determined
from $\chi^2$-error ellipses \citep[e.g.,][]{EISNER+04}.  
We do not include $\beta$ or $\kappa_K$ as free parameters and 
instead consider two model cases:  $\beta=1$ and $\kappa_K=10^3$ (intended
to model sub-micron-sized grains), and $\beta=0$ and $\kappa_K=10^2$ 
(representing grains sized $\sim 10$ $\mu$m).

Scattered light from the disk is justifiably ignored in our modeling.
At $\lambda=1.1$ and 1.6 $\mu$m, the disk-scattered flux on
angular scales of $0\rlap{.}''4$--$4''$ is estimated to comprise
2.4\% and 2.1\% of the stellar flux, respectively \citep{WEINBERGER+02}.  
Given the blue color and roughly flat surface brightness profile
of the scattered light within $\sim 0\rlap{.}''8$ \citep{WEINBERGER+02}, 
we conclude that the $K$-band scattered flux within the 50 mas field of 
view of KI is $< 1\%$.  Therefore thermal emission, as traced by the
$K$-band veiling \citep[$\sim 7\%$ of the stellar flux;][]{JV01},
dominates over any scattered emission.

While we have assumed TW Hya is a single star, a low-mass stellar companion
could contribute to the near-IR visibilities and SED.
To the best of our knowledge, no stellar companions have been detected in 
previous HST imaging or radial velocity monitoring, and thus the presence of 
a luminous second star appears
unlikely. Additional KI observations, or astrometric 
and/or radial velocity monitoring could test this possibility definitively.

\section{Results and Discussion \label{sec:disc}}
\begin{figure}
\plottwo{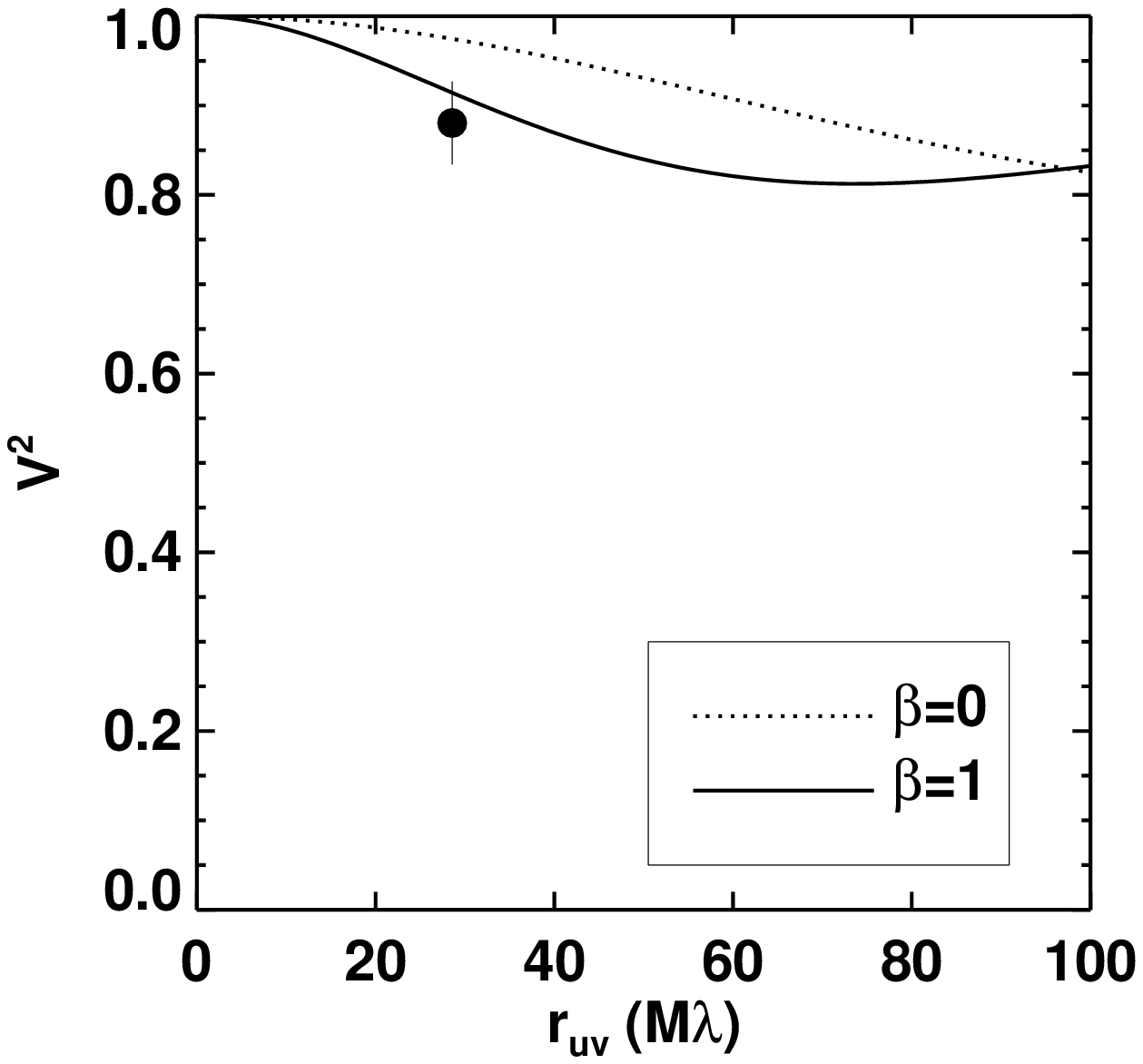}{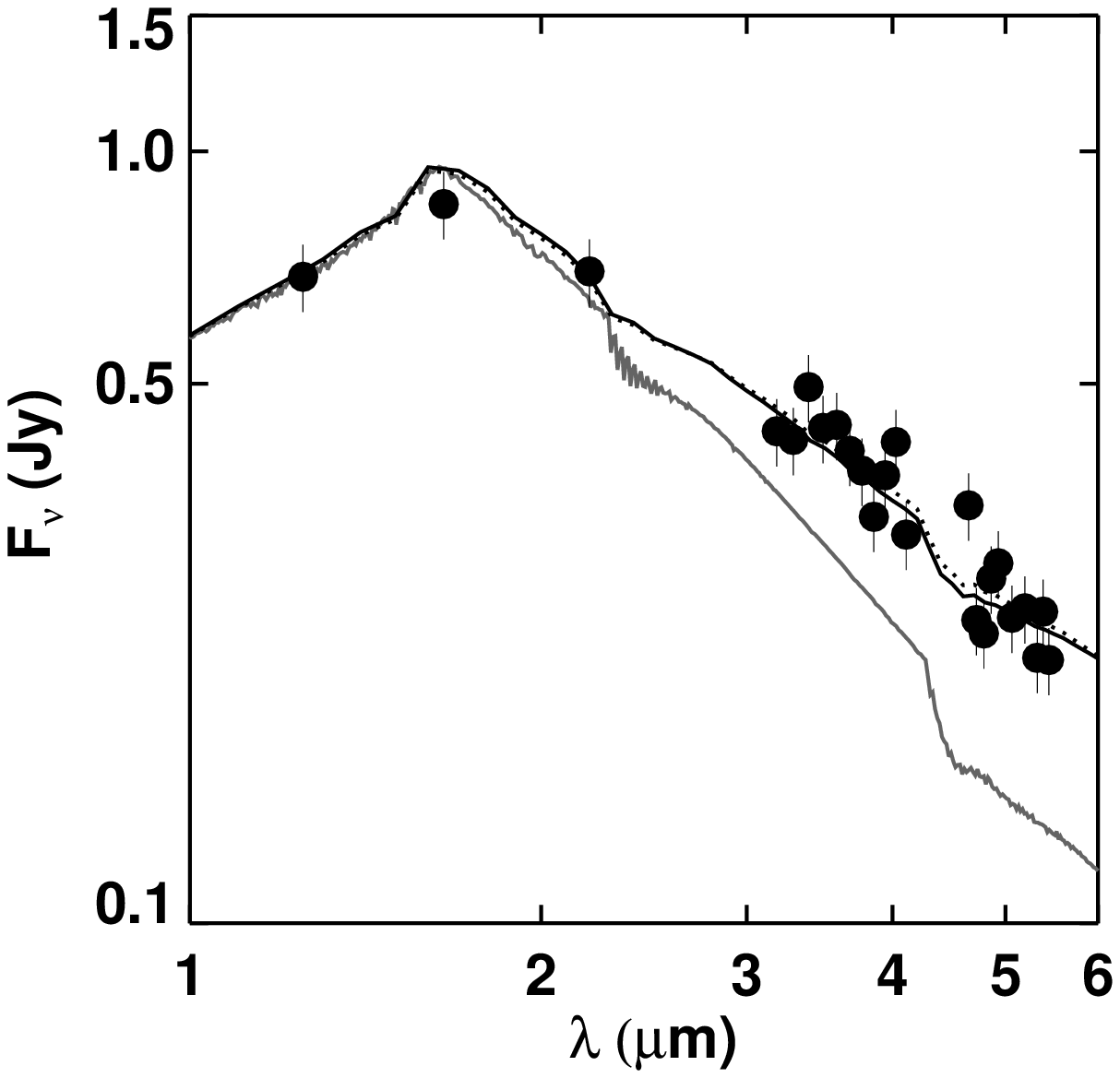}
\caption{Squared visibilities (left) and near-IR
fluxes (right) for a model consisting of a young star surrounded by an 
optically thin disk,
compared to the data \citep[this work;][]{WEBB+99,SLR00}.
The flux of the central star is indicated by the solid gray line.
Models with large dust grains ($\beta=0$) can not fit the
data well, in contrast to models with 
sub-micron-sized dust grains ($\beta = 1$).
\label{fig:mods}}
\end{figure}

Figure \ref{fig:mods} shows the best-fit models for $\beta=0$ and $1$,
together with the $V^2$ and SED data.
Best-fit values for $R_{\rm in}$ and $\Sigma_0$, and their 
$1\sigma$ uncertainties are listed in Table 1.  Our model can 
reproduce the KI $V^2$ measurement and near-IR SED of TW Hya if $\beta= 1$
and $R_{\rm in} \sim 0.06$ AU.  While models with $\beta=0$ can fit the 
SED data well (the larger quantity of SED data relative to the single $V^2$
measurement skews the fits accordingly), only $\beta=1$ models
can simultaneously reproduce our $V^2$ measurement.
From these best-fit parameter values, we compute the temperature at the inner 
truncation radius $T_{\rm in}$, the dust mass $M_{\rm dust}$,\footnote{The
dust mass depends sensitively on $R_{\rm out}$ and the assumed
surface density profile.  Estimates of $M_{\rm dust}$ are therefore 
highly uncertain.  For comparison, under the assumption of a constant surface 
density, \citet{CALVET+02} estimate a dust mass approximately four orders
of magnitude higher than the values listed in Table \ref{tab:models}.} 
and the 2 $\mu$m vertical optical depth at the inner edge $\tau_{K,\rm in}$.

The inner radius of the optically thin disk ($0.06$ AU) exceeds
that inferred from previous modeling of spatially unresolved data 
\citep[0.02 AU;][]{CALVET+02}.
Our large inner radius leads to an inner disk temperature lower than 
expected for dust sublimation (Table 1), suggesting that an alternate 
truncation mechanism is necessary.  One possibility is that the inner disk 
extends inward to the magnetospheric radius $R_{\rm mag}$, 
where the ram pressure from accretion balances the stellar magnetic pressure.  
Although hot dust may still exist interior to 
this radius, its high infall velocity \citep[e.g.,][]{EDWARDS+94} implies an
optical depth orders of magnitude lower than that of dust outside 
$R_{\rm mag}$.  Assuming
an accretion rate of $5 \times 10^{-10}$ M$_{\odot}$ yr$^{-1}$ 
\citep{MUZEROLLE+00}, stellar magnetic field strength of 2.6 kG
\citep{YJV05}, and stellar parameters from \S \ref{sec:mod}, we compute
$R_{\rm mag} \sim 0.09$ AU \citep[e.g.,][]{KONIGL91}.  
This is comparable to our best-fit $R_{\rm in}$ (with $\beta=1$), 
indicating that the optically thin
inner disk may indeed be magnetospherically truncated.  

The fact that models require $\beta = 1$ to fit the combined $V^2$+SED
data indicates that the inner disk contains a population of dust grains 
having sizes smaller than $\sim 1$ $\mu$m.  
In fact, we can obtain slightly better fits to our data if we allow
$\beta>1$, as one might expect if very small grains ($\la
0.01$ $\mu$m) were present \citep{MN93}.  

Sub-micron-sized grains are quickly blown out of the inner disk by stellar
radiation pressure.  Gas friction mediates 
dust removal; unbound grains achieve a terminal outflow velocity equal to the 
product of the momentum stopping time and the net outward acceleration due to 
radiation pressure and gravity 
\citep[e.g.,][]{WEIDENSCHILLING77}.  Estimates of the gas 
density are necessary to calculate
the survival times of small dust grains.  Using the measured column 
density and temperature of H$_2$ in the inner disk \citep{HERCZEG+04},
accounting for the possibility that the midplane may be up to $10^3$ times
denser than the warm surface \citep{GNI04,NAJITA06}, and assuming
a hydrostatic disk, we estimate a midplane gas density at 1 AU 
of $\la 10^{-15}$ 
g cm$^{-3}$.\footnote{Observations of warm CO imply similar gas densities 
\citep{RETTIG+04,NAJITA06}.}
The stopping time for micron-sized grains at $R \sim 1$ AU
is $\ga 10^{-2}$ yr, the terminal velocity is $\ga 1$ km s$^{-1}$, and 
the removal time is $\la 1$ yr.

Small dust grains in the TW Hya inner disk survive for $\la 1$ yr, and 
are thus ephemeral over the age of the system.  Because of the
difficulty in transporting sub-micron-sized grains from the outer disk
at $R \ga 4$ AU to $R_{\rm in}=0.06$ AU, we argue that inner disk dust
is continually re-generated, possibly by collisions of 
a swarm of larger parent bodies that also reside in the inner disk.

\section{Conclusions}
We observed TW Hya with the Keck Interferometer and found the 2 $\mu$m 
emission to be spatially resolved.  We modeled the interferometric 
data together with previous measurements of the $K$-band 
veiling and near-IR fluxes, and inferred that the inner disk consists of 
optically thin dust extending
from the edge of the optically thick outer disk 
\citep[$R \sim 4$ AU;][]{CALVET+02} to $R_{\rm in} = 0.06$ 
AU of the central star.  
This inner radius is larger than expected from dust sublimation; the
truncation may be magnetospheric in origin.  The near-IR
emitting dust is composed of sub-micron-sized particles which are
extremely short-lived; this dust may be
replenished by erosive collisions of larger parent bodies in the inner disk.

\medskip
\noindent{\bf Acknowledgments.}
The near-IR interferometry data presented in this paper were obtained
with the Keck Interferometer (KI) of the W.M. Keck Observatory, which 
was made possible by the generous financial support of the W.M. Keck 
Foundation and is operated as a scientific partnership between the California 
Institute of Technology, the University of California, and NASA.
The authors thank the entire KI team for making these observations possible, 
and acknowledge the cultural role and reverence that the summit of Mauna Kea 
has always had within the indigenous Hawaiian community. 
The authors are grateful to M. Sitko for
providing the 3--5 $\mu$m photometry used in this paper, and to J. Najita and
J. Carr for useful comments about the warm CO gas.  

\epsscale{1.0}

\epsscale{1.0}

\begin{deluxetable}{lccccccc}
\tablewidth{0pt}
\tablecaption{Optically thin disk models \label{tab:models}}
\tablehead{\colhead{$ $} & \colhead{$\chi_{\rm r}^2$} & \colhead{$R_{\rm in}$} 
& \colhead{$R_{\rm out}$} & \colhead{$\Sigma_0$} & \colhead{$T_{\rm in}$} & 
\colhead{$M_{\rm dust}$} &  \colhead{$\tau_{K, \rm in}$}
\\ & & (AU) & (AU) & (g cm$^{-2}$) & (K) & (g) & }
\startdata
$\beta=0$ & 1.17 & $0.02 \pm 0.01$ &  4 & $2.0^{+2.0}_{-0.1} \times 10^{-6}$ & 
1400 & $1 \times 10^{22}$  & 0.07   \\
$\beta=1$ & 0.98 & $0.06 \pm 0.01$ & 4 & $6.3 \pm 1.5 \times 10^{-7}$ & 
1120 & $3 \times 10^{21}$ & 0.04 \\
\enddata
\tablecomments{$R_{\rm in}$ is the best-fit inner radius
and $\Sigma_0$ is the dust surface density at $R=1$ AU.  The outer
disk radius $R_{\rm out}$ is fixed for all models.
The inner disk temperature 
$T_{\rm in}$,  dust mass $M_{\rm dust}$, 
and 2 $\mu$m vertical optical depth at the inner edge $\tau_{K,\rm in}$,
are computed for the best-fit values of $R_{\rm in}$ and $\Sigma_0$.
The value of $M_{\rm dust}$ depends on the assumed surface density profile,
$\Sigma \propto R^{-3/2}$; different assumptions regarding this profile 
yield dust masses that can differ by orders of magnitude.}
\end{deluxetable}

\end{document}